\documentclass[10pt,twocolumn,letterpaper]{article}

\usepackage{iccv}
\usepackage{times}
\usepackage{epsfig}
\usepackage{graphicx}
\usepackage{amsmath}
\usepackage{amssymb}

\usepackage{adjustbox}
\usepackage{gensymb}

\usepackage[breaklinks=true,bookmarks=false]{hyperref}
\hypersetup{
    colorlinks=true,
    linkcolor=blue,
    filecolor=magenta,      
    urlcolor=red,
}
\iccvfinalcopy 

\def\iccvPaperID{****} 

\ificcvfinal\fi

\begin{document}

\title{ AI-MIA: COVID-19 Detection \& Severity Analysis through Medical Imaging}

\author{

Dimitrios Kollias\\
Queen Mary University London, UK\\
{\tt\small d.kollias@qmul.ac.uk} \\

\and 

Anastasios Arsenos \\  
National Technical University\\ Athens, Greece \\

\and
Stefanos Kollias \\ 
National Technical University\\ Athens, Greece     \\

}

\maketitle

\ificcvfinal\thispagestyle{empty}\fi

\begin{abstract}
   
   This paper  presents the baseline approach for the organized 2nd Covid-19 Competition, occurring in the framework of the AIMIA Workshop in the European Conference on Computer Vision (ECCV 2022). It presents the COV19-CT-DB database which is annotated for COVID-19 detction, 
   consisting of about 7,700 3-D CT scans. Part of the database consisting of Covid-19 cases is further annotated in terms of four Covid-19 severity conditions.  We have split the database and the latter part of it in training, validation and test datasets. The former two datasets are used for  training and validation of machine learning models, while the latter will be used for evaluation of the developed models. The baseline approach consists of a deep learning approach, based on a CNN-RNN network and report its performance on the COVID19-CT-DB database.

\end{abstract}

\section{Introduction}

Medical imaging seeks to reveal internal structures of human body, as well as to diagnose and treat diseases. Radiological quantitative image analysis constitutes a significant prognostic factor for diagnosis of diseases, such as lung cancer and Covid-19. For example, CT scan is an essential test for the diagnosis, staging and follow-up of patients suffering from lung cancer in its various histologies. Lung cancer constitutes a top public health issue, with the tendency being towards patient-centered control strategies. Especially in the chest imaging field, there has been a large effort to develop and apply computer-aided diagnosis systems for the detection of lung lesions patterns. 

Deep learning (DL) approaches applied to the quantitative analysis of multimodal medical data offer very promising results. Current targets include building robust and interpretable AI and DL models for analyzing CT scans for the discrimination of immune-inflammatory patterns and their differential diagnosis with SARS-CoV-2-related or other findings. Moreover, detection of the new coronavirus (Covid-19) is of extreme significance for our Society. Currently, the gold standard reverse transcription polymerase chain reaction (RT-PCR) test is identified as the primary diagnostic tool for the Covid-19; however, it is time consuming with a certain number of false positive results. The chest CT has been identified as an important diagnostic tool to complement the RT-PCR test, assisting the diagnosis accuracy and helping people take the right action so as to quickly get appropriate treatment. That is why there is on-going research worldwide to create automated systems that can predict Covid-19 from chest CT scans and  x-rays. 

Recent approaches use Deep Neural Networks for medical imaging problems, including segmentation and automatic detection of the pneumonia region in lungs. The public datasets of RSNA pneumonia detection and LUNA have been used as training data, as well as other Covid-19 datasets. Then a classification network is generally used to discriminate different viral pneumonia \& Covid-19 cases, based on analysis of chest x-rays, or CT scans.  

Various projects are implemented in this framework in Europe, US, China and worldwide. However, a rather fragmented approach is followed: projects are based on specific datasets, selected from smaller, or larger numbers of hospitals, with no proof of good performance generalization over different datasets and clinical environments.  
A DL methodology which we have developed using latent information extraction from trained DNNs has been very successful for diagnosis of Covid-19 through unified chest CT scans and x-rays analysis provided by different medical centers and medical environments.

At the time of CT scan recording, several slices are captured from each person suspected of COVID-19. The large volume of CT scan images calls for a high workload on physicians and radiologists to diagnose COVID-19. Taking this into account and also the rapid increase in number of new and suspected COVID-19 cases, it is evident that there is a need for using machine and deep learning for detecting COVID-19 in CT scans. 

Such approaches require data to be trained on. Therefore, a few databases have been developed consisting of CT scans. However, new data sets with large numbers of 3-D CT scans are needed, so that researchers can train and develop COVID-19 diagnosis systems and trustfully evaluate their performance.    

The current paper presents a baseline approach for the Competition part of the Workshop “AI-enabled Medical Image Analysis – Digital Pathology \& Radiology/COVID19 (AIMIA)” which occurs in conjunction with the European Conference on Computer Vision (ECCV) 2022 in Tel-Aviv, Israel, October 23- 24, 2022. 

The AIMIA-Workshop context is as follows. Deep Learning has made rapid advances in the performance of medical image analysis challenging physicians in their traditional fields. In the pathology and radiology fields, in particular, automated procedures can help to reduce the workload of pathologists and radiologists and increase the accuracy and precision of medical image assessment, which is often considered subjective and not optimally reproducible. In addition, Deep Learning and Computer Vision demonstrate the ability/potential to extract more clinically relevant information from medical images than what is possible in current routine clinical practice by human assessors. Nevertheless, considerable development and validation work lie ahead before AI-based methods can be fully ready for integrated into medical departments.

The workshop on AI-enabled medical image analysis (AIMIA) at ECCV 2022 aims to foster discussion and presentation of ideas to tackle the challenges of whole slide image and CT/MRI/X-ray analysis/processing and identify research opportunities in the context of Digital Pathology and Radiology/COVID19.

The COV19D Competition is based on a new large database of chest CT scan series that is manually annotated for Covid-19/non-Covid-19, as well as Covid-19 severity ndiagnosis. The training and validation partitions along with their annotations are provided to the participating teams to develop AI/ML/DL models for Covid-19/non-Covid-19 and Covid-19 severity prediction. Performance of approaches will be evaluated on the test sets.

The COV19-CT-DB is a new large database with about 7,700 3-D CT scans, annotated for COVID-19 infection. 

The rest of the paper is organized  as follows. Section 2 presents former work on which the presented baseline has been based. Section 3 presents the database created and used in the Competition. The ML approach and the pre-processing steps  are described in Section 4. The obtained results, are presented in Section 5. Conclusions and future work are described in Section 6.

\section{Related Work}

In \cite{khadidos2020analysis} a CNN plus RNN network was used, taking as input CT scan images and discriminating
between COVID-19 and non-COVID-19 cases. 

In \cite{li2020coronavirus}, the authors employed a variety of 3-D ResNet models for detecting COVID-19 and distinguishing it from other common pneumonia (CP) and normal cases, using  volumetric 3-D CT scans. 

In \cite{wang2020weakly}, a weakly supervised deep learning framework was suggested using 3-D CT volumes for COVID-19 classification and lesion localization. A pre-trained UNet was utilized for segmenting the lung region of each CT scan slice; the latter was fed into a 3-D DNN that provided the classification outputs.  

The presented approach is based on a CNN-RNN architecture that performs 3-D CT scan analysis. The method follows our previous work \cite{Tailor, springer, cis, ijait} on developing deep neural architectures for  predicting COVID-19, as well as neurodegenerative and other \cite{iet, cis, 48aposcholarmou, mdpi} diseases and abnormal situations \cite{art11, art12, art13, art14}.

These architectures have been applied for: a) prediction of Parkinson’s, based on datasets of MRI and DaTScans, either created in collaboration with the Georgios Gennimatas Hospital (GGH) in Athens \cite{cis}, or provided by the PPMI study sponsored by M. J. Fox for Parkinson’s Research \cite{iet}, b) prediction of COVID-19, based on CT chest scans, scan series, or x-rays, either collected from the public domain, or aggregated in collaboration with the Hellenic Ministry of Health and The Greek National Infrastructures for Research and Technology \cite{Tailor}.

The current Competition follows and extends the MIA-COV19D Workshop we successfully organized virtually in ICCV 2021, on October 11, 2021. 35 Teams participated in the COV19D Competition; 18 Teams submitted their results and 12 Teams scored higher than the baseline. 21 presentations were made in the Workshop, including 3 invited talks \cite{ref99}. 

In general, some workshops on medical image analysis have taken place in the framework of CVPR/ICCV/ECCV Conferences in the last three years. In particular: 

In ICCV 2019 the 'Visual Recognition for Medical Images' Workshop took place, including papers on the use of deep learning in medical images for diagnosis of diseases. Similarly, in CVPR 2019 and 2020 the following series of Workshops took place: 'Medical Computer Vision' and 'Computer Vision for Microscopy Image Analysis', dealing with general medical analysis topics. In addition, the ‘Bioimage Computing' took place in ECCV 2020. 
Moreover, the issue of explainable DL and AI has been a topic of other Workshops (e.g., CVPR 2019; ICPR 2020, \& 2022; ECAI Tailor 2020)

\section{The COV19-CT-DB Database}

The COVID19-CT-Database (COV19-CT-DB) consists of chest CT scans that are annotated for the existence of COVID-19. 

COV19-CT-DB includes 3-D chest CT scans  annotated for existence of COVID-19. Data collection was conducted in the period from September 1 2020 to November 30 2021.  It consists of 1,650 COVID and 6,100 non-COVID chest CT scan series, which correspond to a high number of patients (more than 1150) and subjects (more than 2600). In total, 724,273 slices correspond to the CT scans of the COVID-19 category and 1,775,727 slices correspond to the non COVID-19 category. 

Annotation of each CT slice has been performed by 4 very experienced (each with over 20 years of experience) medical experts; two radiologists and two pulmonologists. Labels provided by the 4 experts showed a high degree of agreement (around 98\%). Each of the 3-D scans includes different number of slices, ranging from 50 to 700. This variation in number of slices
is due to context of CT scanning. The context is defined in terms of various factors, such as the accuracy asked by the doctor who ordered the scan, the characteristics of the CT scanner that is used, or specific subject’s features, e.g., weight and age.

\begin{figure*}[h!]
\centering
\includegraphics[height=0.19\linewidth]{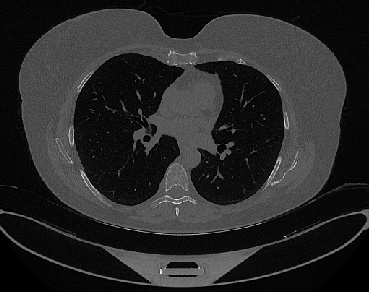}
\includegraphics[height=0.19\linewidth]{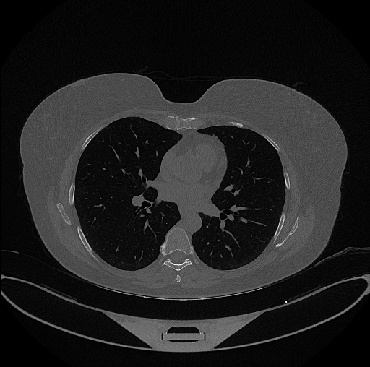}
\includegraphics[height=0.19\linewidth]{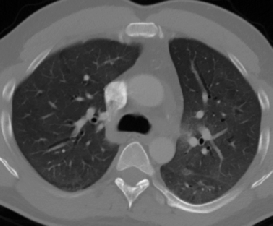}
\includegraphics[height=0.19\linewidth]{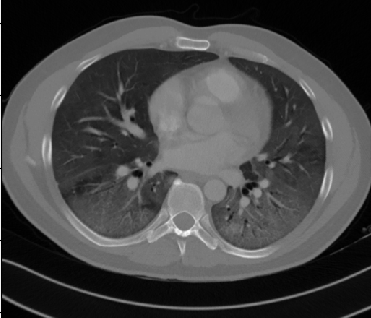}
\caption{Four CT scan slices, two from a non-COVID-19 CT scan, on the left and two from a COVID-19 scan, on the right, including bilateral ground glass regions in lower lung lobes.} 
\label{samples}
\end{figure*}

Figure \ref{samples} shows four CT scan slices, two from a non-COVID-19 CT scan, on the left and two from a COVID-19 scan, on the right. Bilateral ground glass regions are seen especially in lower lung lobes in the COVID-19 slices. 

The database has been split in training, validation and testing sets. 

The training set contains, in total, 1992 3-D CT scans.  The validation set consists of 494 3-D CT scans.  The number of  COVID-19 and of  Non-COVID-19 cases in each set are shown in Table \ref{splits_ch1}. There are different numbers of CT slices per CT scan, ranging from 50 to 700.  

\begin{table}[h]
\caption{Data samples in each Set }
\label{splits_ch1}
\centering
\scalebox{0.7}{
\begin{tabular}{| c | c| c |}

\hline Set & Training & Validation\\
\hline
\hline
COVID-19   & 882 & 215 \\
\hline
Non-COVID-19 & 1110 & 289 \\
\hline
\end{tabular}
}
\end{table}

A further split of the COVID-19 cases has been implemented, based on the severity of COVID-19, as annotated by four medical experts, in the range from 1 to 4, with 4 denoting the critical status. Table \ref{tablecategories} describes each of these categories \cite{ref13}.  

\begin{table}[h]
\caption{Description of the Severity Categories }
\label{tablecategories}
\centering
\scalebox{0.7}{
\begin{tabular}{| c | c | p{0.8\linewidth}|}

\hline
Category & Severity & Description\\
\hline
\hline
1 & Mild   & Few or no Ground glass opacities. Pulmonary parenchymal involvement $\leq 25 \%$ or absence \\
\hline
2 & Moderate  & Ground glass opacities.  Pulmonary parenchymal involvement  $25-50$\% \\
\hline
3 & Severe & Ground glass opacities. Pulmonary parenchymal involvement  $50-75$\% \\
\hline
4 & Critical &   Ground glass opacities. Pulmonary parenchymal involvement  $\geq 75$\%  \\
\hline
\end{tabular}
}
\end{table}

In particular, parts of the COV19-CT-DB COVID-19 training and validation datasets have been accordingly split for severity classification in training, validation and testing sets. 

The training set contains, in total, 258 3-D CT scans.  The validation set consists of 61 3-D CT scans.  The numbers of scans in each severity COVID-19 class in these sets  are shown in Table \ref{splits_ch2}.

\begin{table}[h]
\caption{Data samples in each Severity Class }
\label{splits_ch2}
\centering
\scalebox{0.7}{
\begin{tabular}{| c | c| c |}

\hline Severity Class & Training & Validation\\
\hline
\hline
1  & 85 & 22 \\
\hline
2 & 62 & 10 \\
\hline
3  & 85 & 22 \\
\hline
3 & 26 & 5 \\
\hline
\end{tabular}
}
\end{table}

\section{The Deep Learning Approach}

\subsection{3-D Analysis and COVID-19 Diagnosis}

The input sequence is a 3-D signal, consisting of a series of chest CT slices, i.e., 2-D images, the number of which is varying, depending on the context of CT scanning. The context is defined in terms of various requirements, such as the accuracy asked by the doctor who ordered the scan, the characteristics of the CT scanner that is used, or the specific subject’s features, e.g., weight and age.

\begin{figure*}[h!]
\centering
\adjincludegraphics[height=11cm]{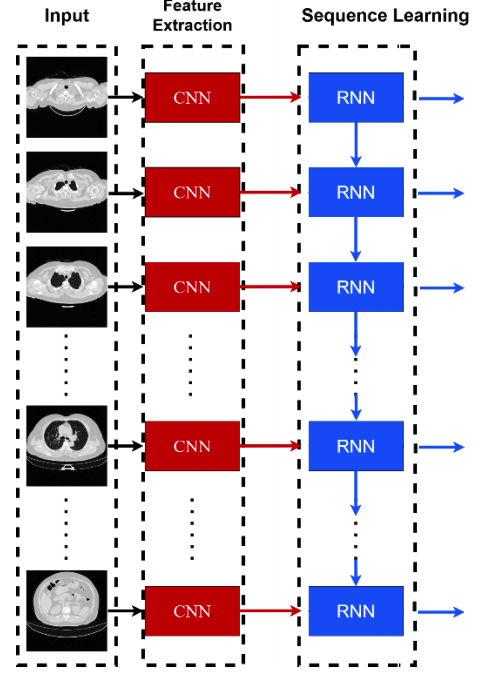}
\caption{The CNN-RNN model}
\label{cnn_rnn}
\end{figure*}

The baseline approach is a CNN-RNN architecture, as shown in Figure \ref{cnn_rnn}. At first all input CT scans are padded to have length $t$ (i.e., consist of $t$ slices). The whole (unsegmented) sequence of 2-D slices of a CT-scan are fed as input to the CNN part. Thus the CNN part performs local, per 2-D slice, analysis, extracting features mainly from the lung regions. The target is to make diagnosis using the whole 3-D CT scan series, similarly to the annotations provided by the medical experts. The RNN part provides this decision, analyzing the CNN features of the whole 3-D CT scan, sequentially moving from slice $0$ to slice $t-1$. The outputs of the RNN part feed the output layer -with 2 units- that uses a softmax activation function and provides the final COVID-19 diagnosis. 

In this way, the CNN-RNN network outputs a probability for each CT scan slice; the CNN-RNN is followed by a voting scheme that makes the final decision; the voting scheme can be either a majority voting or an at-least one voting (i.e., if at least one slice in the scan is predicted as COVID-19, then the whole CT scan is diagnosed as COVID-19, and if all slices in the scan are predicted as non-COVID-19, then the whole CT scan is diagnosed as non-COVID-19).

\subsection{Pre-Processing \&  Implementation Details}

At first, CT images were extracted from DICOM files. Then, the voxel intensity values were clipped using a window/level of $350$ Hounsfield units (HU)/$-1150$ HU and normalized to the range of $[0, 1]$. 

Regarding implementation of the proposed methodology: i) we utilized ResNet50 as CNN model,  stacking on top of it a global average pooling layer, a batch normalization layer and dropout (with keep probability 0.8); ii) we used a single one-directional GRU layer consisting of 128 units as RNN model. The model was fed with 3-D CT scans composed of the CT slices; each slice was resized from its original size of $512 \times 512 \times 3$ to  $224 \times 224 \times 3$. As a voting scheme, we used the at-least one.

Batch size was equal to 5 (i.e, at each iteration our model processed 5 CT scans) and the input length 't' was 700 (the maximum number of slices found across all CT scans). Softmax cross entropy was the utilized loss function for training the model. Adam optimizer was used with learning rate   $10^{-4}$. Training was performed on a Tesla V100 32GB GPU.

The architecture trained for Challenge 1 was then refined to tackle the severity classification problem of Challenge 2. 

\section{Experimental Results}

This section describes a set of experiments evaluating the performance  of the baseline approach.

Table \ref{3dcnn_rnn} shows the performance of the network over the validation sets in both Challenges, after training with the training datasets, in terms of macro F1 score. The macro F1 score is defined as the unweighted average of the class-wise/label-wise F1-scores, i.e., the unweighted average of the COVID-19 class F1 score and of the non-COVID-19 class F1 score.

The main downside of the model is that there exists only one label for the whole CT scan and there are no labels for each CT scan slice. Thus, the presented model analyzes the whole CT scan, based on information extracted from each slice.

\begin{table}[t]
\caption{Performance of the baseline ResNet50-GRU network for the two Challenges}
\label{3dcnn_rnn}
\centering
\scalebox{1.}{
\begin{tabular}{|c|c|}
\hline
 Challenge  &  \multicolumn{1}{c|}{'macro' F1 Score}\\
 \hline
 \hline
 COVID-19 Detection   &   0.77  \\
\hline
 COVID-19 Severity Detection   &   0.63  \\
\hline
\end{tabular}
}
\end{table}

\section{Conclusions and Future Work}

In this paper we have introduced  a new large database of chest 3-D CT  scans, obtained in various contexts and consisting of  
different numbers of CT slices. We have also developed a deep neural network, based on a CNN-RNN architecture and used it for  COVID-19 diagnosis on this database. 


The scope of the paper is to present a baseline scheme regarding the performance that can be achieved based on analysis of the COV19-CT-DB database.  

The model presented in the paper will be the basis for future expansion towards more transparent modelling of COVID-19 diagnosis.    

{\small
\bibliographystyle{ieee_fullname}
\bibliography{egbib}
}

\end{document}